\documentstyle[preprint,aps]{revtex}
\begin{document}
\title{A new orthogonalization procedure with an extremal property} 
\author{S. Chaturvedi\thanks{e-mail:scsp@uohyd.ernet.in}, A.K. Kapoor
\thanks{email:akksp@uohyd.ernet.in} and V. Srinivasan
\thanks {e-mail:vssp@uohyd.ernet.in}}
\address{School of Physics\\
University of Hyderabad\\ Hyderabad - 500 046 (INDIA)}
\maketitle
\begin{abstract}
Various methods of constructing an orthonomal set out of a given set 
of linearly independent vectors are discussed. Particular attention 
is paid to the Gram-Schmidt and the Schweinler-Wigner orthogonalization 
procedures. A new orthogonalization procedure which, like the Schweinler-
Wigner procedure, is democratic and is endowed with an extremal property 
is suggested. 
\end{abstract}
\newpage
Constructing an orthonormal set out of a given set of linearly independent 
Vectors is an age old problem. Among the many possible orthogonalization 
procedures, the two algorithmic procedures that have been extensively 
discussed and used in the literature are (a) the familiar Gram-Schmidt 
procedure [1]  and (b) a procedure which is referred to as the Schweinler
-Wigner procedure [2], particularly in the wavelet literature [3]. ( This 
method is known among the chemists as the L\"owdin orthogonalization 
procedure [4]. Mathematicians attribute it to Poincar\'e. Schweinler and 
Wigner themselves trace its origin to a work of Landshoff [5]. Eschewing the 
question of historically correct attribution, we shall continue to refer to 
it as the Schweinler-Wigner procedure ) An intrinsic difference between the 
two procedures is that while  the Gram-Schmidt procedure, by its very nature, 
requires one to select the linearly independent vectors sequentially,   
the Schweinler-Wigner procedures treats all the members of the set of 
linearly independent vectors democratically. The significance of the work 
of Schweinler and Wigner lies not in  introducing a new orthogonalization 
method - the method was already known but rather in introducing a positive 
quantity $m$, to be defined shortly, which discriminates between various 
orthogonalization 
procedures. They showed that $m$ is a maximum for the Schweinler-Wigner 
basis. In this letter we pose and answer the question as to what is the 
orthogonalization procedure which minimizes $m$. This new orthogonalization 
procedure, like the Schweinler-Wigner procedure, also turns out to be 
completely democratic in that it treats all the linearly independent 
vectors on the same footing. The quantity $m$ was introduced 
by Schweinler and Wigner in a some what ad-hoc manner. We reformulate 
their procedure in a way so as that the quantity $m$ appears in a natural 
way and can be useful in a wider context than that for which it was 
introduced. In particular, this reformulation enables us to quantify 
the notion of an orthonormal basis which brings any Hermitian operator 
in to a maximally off-diagonal form.      
      
Let $v_1,\cdots,v_N$ denote a set of $N$ linearly independent vectors. Let 
$M$ denote the associated Gram matrix :$ M_{ij}= (v_i,v_j)$. $M$ is a positive 
definite Hermitian matrix.
Define 
\begin{equation}
{\bf z}= {\bf v}S ~~~,
\end{equation}
where S is an invertible matrix. Then 
\begin{equation}
(z_i,z_j) = (S^\dagger M S)_{ij}~~~.
\end{equation}
Requiring that ${\bf z}$ be an orthonormal basis amounts to requiring that  
\begin{equation} 
S^\dagger M S = I~~~ i.e.~~~~ M^{-1} = SS^{\dagger}~~~.
\end{equation}
Each such $S$ defines an orthogonalization procedure. Two standard choices of 
$S$ are 
\\
\noindent
[1] Schweinler-Wigner Procedure

This procedure corresponds to the choice
\begin{equation}
S = U P^{-1/2} U^{\dagger}~~~,
\end{equation}
where $U$ is the matrix which brings $M$ to a diagonal form $P$
\begin{equation}
U^\dagger M U = P~~~.
\end{equation}
With this choice of $S$, which corresponds to taking the Hermitian square root 
of the matrix $M$, one has 
\begin{equation}
{\bf z} = {\bf v} U P^{-1/2}U^{\dagger}~~~.
\end{equation}
On defining $ {\bf w } = {\bf z} U $, one obtains the Schweinler-Wigner 
basis 
\begin{equation}
{\bf w} = {\bf v} U P^{-1/2} ~~~.
\end{equation}

\noindent
[2] Gram-Schmidt Orthogonalization Procedure

In this procedure $S$ is chosen to be an upper triangular matrix $T$
 satisfying 
\begin{equation}
M^{-1} = TT^\dagger
\end{equation}
and the Gram-Schmidt basis is given by
\begin{equation}
{\bf y} = {\bf v} T
\end{equation}

The two  orthonormal bases ${\bf w}$ and ${\bf y}$ discussed above are related 
to each other by the following unitary transformation 
\begin{equation}
{\bf y} = {\bf w} V^{(1)}~~~,
\end{equation}
where $V^{(1)}$ is given by
\begin{equation}
V^{(1)} = P^{1/2}U^{\dagger}T~~~.
\end{equation} 
Schweinler and Wigner introduced a quantity $m({\bf z})$ as follows
\begin{equation}
m({\bf z})=\sum_{k}\left(\sum_{l} {\mid (z_k,v_l)\mid }^2 \right)^2~~~,
\end{equation}
where ${\bf z}$ is any orthonormal basis. They further showed that 
$m({\bf z})$ attains its maximum value $Tr(M^{2})$ for ${\bf z} ={\bf w}$. 
\begin{equation}
m_{max} = Tr(M^{2}) = m({\bf w})~~~.
\end{equation} 
For any other basis ${\bf z}$, related to ${\bf w}$ by a unitary transformation
V
\begin{equation}
{\bf z} = {\bf w} V~~~,
\end{equation}
the value of $m({\bf z})$ is given by 
\begin{equation}
m({\bf z})=\sum_{k}\left({\left(V^\dagger P V\right)}_{kk}\right)^2~~~.
\end{equation}
In particular, for the Gram-Schmidt basis ${\bf y}$, one finds that
\begin{equation}
m({\bf y}) =\sum_{k} \left({{\left(T^\dagger T\right)}^{-1}}_{kk} \right)^2 
~~~. 
\end{equation}

A natural question to ask is as to what is the  orthonormal basis  
which minimizes $m({\bf z})$. On applying Cauchy inequality to $(15)$, 
it follows that 
\begin{eqnarray}
m({\bf z}) &=&\sum_{k}\left({\left(V^\dagger P V\right)}_{kk}\right)^2 
\nonumber \\
&\geq& \frac{1}{N} \left( \sum_{k} (V^\dagger P V)_{kk}\right)^2 \nonumber \\
&\geq& \frac{1}{N} (Tr M)^2~~~. 
\end{eqnarray}
Equality sign holds if and only if $(V^\dagger P V)_{kk} = c$ independent of 
$k$ i.e 
\begin{equation}
\sum_{l} P_l \mid V_{kl}\mid^2 = \frac{1}{N} (P_1+\cdots+P_N)~~~
\end{equation}
This requires that 
\begin{equation}
\mid V_{kl}\mid^2 = \frac{1}{N}~~~, 
\end{equation}
for all $k$ and $l$. 
The matrix elements of the unitary matrix $V^{(2)}$ satisfying the above 
equation are thus given by 
\begin{equation}
V^{(2)}_{kl} = \frac{1}{\sqrt N}\exp{\left[\frac{2\pi i (k-1)(l-1)}{N}\right]}
~~~.
\end{equation}
The matrix $V^{(2)}$ is thus just the character table of the cyclic group 
$C_N$. Thus the basis ${\bf x}$, for which $m({\bf z})$ attains its minimum 
value $m_{min} = (1/N)(TrM)^2$, is related to the Schweinler-Wigner basis 
${\bf w}$ as follows 
\begin{equation}
 x_l =\frac{1}{\sqrt N} \sum_{k} \exp{\left[\frac{2\pi i (k-1)(l-1)}{N}\right]}
 w_k ~~~.
\end{equation}
This basis also treats all the linearly independent vectors ${\bf v}$ 
democratically like the Schweinler-Wigner basis.

The quantity $m({\bf z})$ appears to have been introduced by Schweinler 
and Wigner in a rather ad-hoc way. At least no particular motivation 
for introducing it appears in their work. We now reformulate their work 
in a way that this quantity appears naturally. Consider the 
Hermitian operator 
\begin{equation}
{\cal M} = \sum_{k} v_{k} v_{k}^{\dagger}~~~. 
\end{equation}
In an arbitrary orthonormal basis ${\bf z}$ one can write 
\begin{equation}
Tr({\cal M}^2) = \sum_{lm \atop{l\neq m}} {\mid (z_l,{\cal M} z_m)\mid}^2  
+ \sum_l {\mid (z_l,{\cal M} z_l)\mid}^2~~~.  
\end{equation} 
The second term on the rhs is easily seen to be the same as $m({\bf z})$ 
in $(12)$. From this perspective it is immediately obvious that the basis 
which maximizes $m({\bf z})$ is the one in which ${\cal M}$ is diagonal. 
This is just the Schweinler-Wigner basis as can also be directly verified. 
Thus the Schweinler-Wigner basis is simply the eigenbasis of the operator 
${\cal M}$ and if the eigenvalues of ${\cal M}$ are all distinct then this 
basis is essentially unique. Further, since $Tr({\cal M}^2)$ is independent 
of the choice of basis, it is clear from $(23)$ that the basis which 
minimizes $m({\bf z})$ maximizes
\begin{equation}
n({\bf z}) \equiv \sum_{lm \atop{l\neq m}} {\mid (z_l,{\cal M} z_m)\mid}^2~~.
\end{equation}
The quantity $n({\bf z})$ therefore provides a quantitative measure of the 
off-diagonality of the operator ${\cal M}$ in the ${\bf z}$ basis. 
The new orthonormal basis proposed in this work is thus the one 
in which ${\cal M}$ is maximally off-diagonal.

\vskip3.0cm 
\noindent{\bf Acknowledgements}

We are grateful to Prof G.S. Agarwal, Prof Vipin Srivastava and to Dr. 
M. Durgaprasad for numerous discussions.

\newpage
\noindent{\bf References}
\begin{enumerate}
\item[[1]] See, for instance, Gantmacher F C 1960 {\it The Theory of Matrices}
 Vol 1  (New York : Chelsea)
\item[[2]] Schweinler H C  and Wigner E P 1970  J. Math.  Phys. {\bf 11} 1693
\item[[3]] See, for instance, Chui C K 1992  {\it Wavelet Analysis and its 
Applications}  (San Diego CA : Academic Press)
\item[[4]] L\"owdin P O 1950 J. Chem. Phys. {\bf18}  365 ; 1956 Adv. Phys. 
{\bf 5}  172 ; 1992 Adv. Quant. Chem. {\bf 23}, 84  
\item[[5]] Landshoff R  1936 Z. Physik {\bf 102}, 201 
\end{enumerate}
\end{document}